\documentclass[runningheads]{svmult}

\usepackage{makeidx}   
\usepackage{graphicx}  
\usepackage{subeqnar}  
\usepackage{multicol}  
\usepackage{physprbb}  
\makeindex             

\usepackage{amssymb}
\usepackage{color}
\newcommand{\Msun}{\rm{M}_{\odot}}

\begin{document}
\title*{Gamma-Ray Bursts and Jet-Powered Supernovae}
\toctitle{Gamma-Ray Bursts and Jet-Powered Supernovae}
%
%
\titlerunning{GRBs and Jet-Powered Supernovae}
%
\author{S. E. Woosley\inst{1}
Weiqun Zhang\inst{1}
\and A. Heger\inst{2}
}
\authorrunning{Woosley et al.}
%
%
\institute{
Department of Astronomy and Astrophysics,
University of California, \\
Santa Cruz, CA 95064, U.S.A.
\and 
Department of Astronomy and Astrophysics, 
Enrico Fermi Institute,\\
The University of Chicago,
5640 S.\ Ellis Ave,
Chicago, IL 60637,
U.S.A.
}

\maketitle              

\begin{abstract}

The last five years have seen growing challenges to the traditional
paradigm of a core collapse supernova powered by the neutrino emission
of a young proto-neutron star. Chief among these challenges are
gamma-ray bursts (GRBs) and the supernovae that seem to accompany
them.  Here we review some recent - and not so recent - models for
GRBs and supernovae in which strong magnetic fields, rotation, or
accretion into a black hole play a role. The conditions for these
energetic explosions are special and, at this point, there is no
compelling reason to invoke them in the general case. That is, 99\% of
supernovae may still operate in the traditional fashion.
\end{abstract}

\section{Introduction}

The demise of spherically symmetric models for supernovae can be
traced to 1987. Though we certainly already understood that stars (and
pulsars) rotated and had magnetic fields and that this might affect
the explosion \cite{Leb70}, that neutrino powered convection had to be
included in any realistic model \cite{wil86}, and that instabilities
would be encountered as the shock moved out \cite{CK78}, it was the
clear evidence for mixing on a large scale in SN 1987A \cite{87A} that
drove us inexorably to multi-dimensional models. The migration was
facilitated by developments in computer hardware and software that
made multi-dimensional calculations practical. Still hope remained
that, globally, things would still be pretty spherically symmetric. In
particular, the shock wave coming out from the neutron star, though
bounding regions that bubbled and mixed, was roughly spherical.

Events of similar significance happened in 1997, when it became
clear that GRBs are located at cosmological distances
\cite{cos97,ian97}, and again in 1998 when a supernova, SN 1998bw, was
discovered in conjunction with a GRB. The supernova had very peculiar
properties and, if modeled in one dimension (surely a gross
approximation), had a kinetic energy in excess of 10$^{52}$ erg
\cite{Woo98,iwa98}.  Even more dramatic was the discovery that a
significant fraction of that energy was contained in relativistic
ejecta \cite{kul98}.

During the next four years evidence accumulated both for supernovae
associated with GRBs \cite{Blo99,Rei99,Blo02a} and for
unusually energetic supernovae (see talk by Nomoto). The term
``hypernovae'' \cite{pac98} was often used to describe these
exceptional explosions, and has lately come to denote almost any
unusual supernova with inferred high energy (along the line of sight)
or broad lines. Here we will avoid the term, which is not associated
with any particular model, and speak of the specific {\sl mechanisms}
that might be responsible for exploding stars with great energy, gross
asymmetry, and/or relativistic mass ejection. Obviously
energy and asymmetry are not independent. A grossly asymmetric
explosion may appear anomalously energetic - in terms of broad lines
for example - along one line of sight and not another.

In this regard, GRBs themselves may be an extreme case of a continuous
distribution of events ranging from nearly spherical supernovae with
kinetic energies of order 10$^{51}$ erg, to events like GRB 990123 with
an inferred equivalent isotropic energy of over $10^{54}$ erg. But is
it energy or asymmetry? Recent analysis of the afterglows of GRBs
\cite{Fra01} has shown that the total energies in GRBs are really
remarkably clustered around 10$^{51}$ erg, even for 990123, and that
their exceptional brilliance is a consequence of having focused some
appreciable fraction of that energy into a narrow, relativistic jet
($\Gamma \sim 200$) moving in our direction. Other observations have
also shown the association of GRBs with star forming regions inside
galaxies \cite{Blo02b}. Taken together, a picture is emerging that at
least some massive stars die while producing relativistic jets.

\section{Jet-Powered Supernovae (JetSN) and Pulsar-Powered Supernovae}

\subsection{Rotation}

All modern models for GRBs and JetSN invoke rapid rotation, either of
a neutron star or of a disk around a black hole. For typical equations
of state, a neutron star with radius 10 km and period $\sim$5 ms has
$\sim$10$^{51}$ erg of rotational kinetic energy, and this is an upper
bound on the final period that is needed, especially since most of the
action occurs when the radius is 30 km, not 10 km.

The evolution of massive stars including the transport of angular
momentum by magnetic \cite{Heg00} and non-magnetic processes
\cite{Heg02a} has been considered until core collapse in various
papers by Heger, Woosley, Spruit, and Langer. To summarize, common red
supergiants, the progenitors of most supernovae, end up producing
neutron stars with rotation rates near break up when magnetic fields
are omitted, and around 10 ms when current estimates of magnetic
torques \cite{Spr01} {\sl are} included. The 10 ms value accounts for
angular momentum loss due to neutrinos flowing out of the neutron
star, but does not include possible braking by a neutrino-powered
magnetic stellar wind or by the propeller mechanism operating in
conjunction with fallback \cite{Heg02a}.

For GRB progenitors, a bare helium core is more appropriate.  A helium
star born (e.g., from a merger) with equatorial rotation 10\% of
Keplerian and low metallicity can retain enough angular momentum to
form a centrifugally supported disk around a central (Kerr) black hole
of $\sim3\,\Msun$ provided that magnetic fields are left out of the
calculation \cite{Heg00,Heg02b}.  However, when an approximate
treatment of angular momentum transport by magnetic fields is included
\cite{Spr01} along with mass loss, the resulting rotation become too
low to form centrifugally supported disks in the inner part of the
core \cite{Heg02b}.

Admittedly our knowledge of magnetic torques inside evolved massive
stars is still uncertain, but these results suggest that: a) magnetic
field torques during the pre-supernova evolution are an important
consideration, and b) within uncertainties, all current models may be
allowed, but may require special circumstances. This may be why GRBs
only occur in about 1\% of supernovae (based on estimates of GRB
beaming and the supernova rate in the universe).  

Since nature is continuous, however, we may also expect many
supernovae in which rotation plays an important role (i.e., the
inferred pulsar rotation rate is faster than 5 ms), but no GRB is
produced.

\subsection{Pulsar-powered supernovae}

Shortly after pulsars were discovered and their rapid rotation rates
inferred, it was suggested that they might power supernovae
\cite{Ost71}. If energies of $> 10^{50}$ erg must be rapidly
dissipated by means other than neutrinos or gravity waves, it is
unavoidable that a pulsar will influence supernova dynamics, leading,
for example, to additional mixing. However, pulsars as the cause of
common supernova explosions encounters at least two objections. First,
the accretion rate shortly after neutron star formation is $\sim$0.1
to 1 $\Msun$ s$^{-1}$. The Alfven radius for this accretion rate is
then \cite{Alp01}
\begin{equation}
r_A = 1.3 \times 10^4 \ {\rm cm} \ \mu_{30}^{4/7} \, \dot M_{32}^{-2/7}
\end{equation}
with $\mu_{30}$ the magnetic moment in G cm$^{-3}$ (10$^{30}$ is
approximately the value for B $\sim$ 10$^{12}$ G) and $\dot M_{32}$,
the accretion rate in units 10$^{32}$ g s$^{-1}$. For the Alfven
radius to be greater than the neutron star radius, $\sim$10 km, with
an accretion rate of 0.3 $\Msun$ s$^{-1}$ the magnetic moment must
exceed $5 \times 10^{33}$ and the B field must exceed $5 \times
10^{15}$ G. When the explosion is developing, the protoneutron star
radius is actually more like 30 km and the necessary magnetic moment
about 10 times greater.  This implies, baring ultrastrong magnetic
fields, that no pulsar will be able to function during the critical
epoch when the accretion rate is high and the probability of black
hole formation large.

Second is the issue of $^{56}$Ni nucleosynthesis. A shock like the one
produced in neutrino powered explosions will raise a significant
quantity of material to temperatures greater than $5 \times 10^9$ K
and thus make iron group elements \cite{WHW02}. To do so the shock
must receive its energy in a time short compared with that
needed to cross the region where the nickel is made, about 4000 km.
This is, at most, a few tenths of a second. If a pulsar does not
deposit at least 10$^{51}$ erg in this brief interval, very little
nickel will be made to power the light curve. Such short braking times
again require very large magnetic fields and rotation rates. While it
may be that the occasional neutron star is born with these extreme
properties (see below), we do not think it happens in most supernovae.

\begin{figure}[t]
\begin{center}
\includegraphics[width=.8\textwidth]{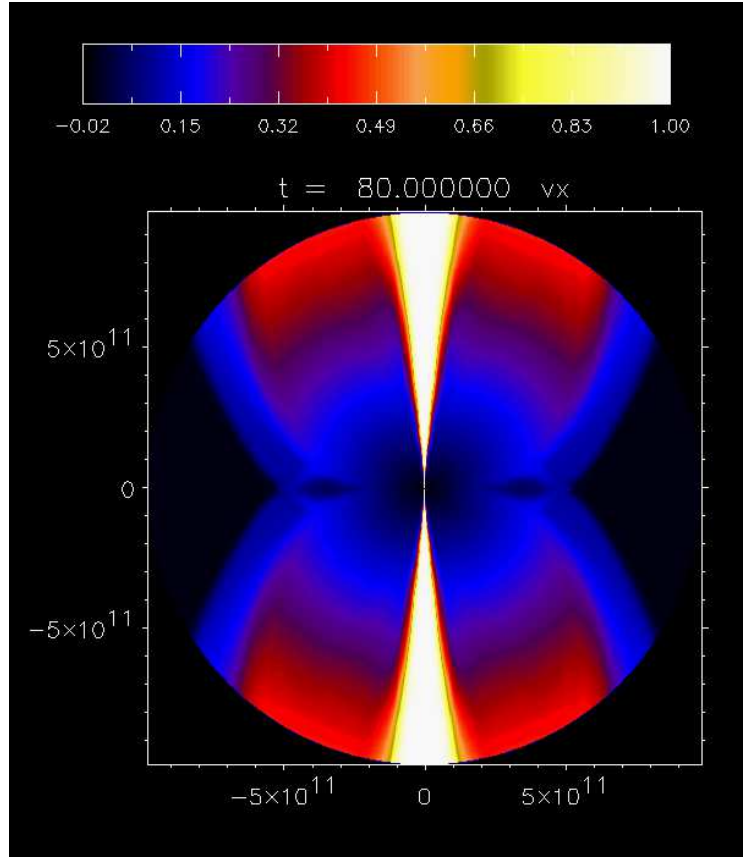}
\end{center}
\caption[]{A massive Wolf-Rayet star being exploded by the passage of
relativistic jets along its axes \cite{Zha02b}.  The jet was initiated
at 2000 km in a Wolf-Rayet star with radius 700,000 km and had a
Lorentz factor of 10 for the first 10 seconds which slowly declined to
2 at 1000 s. The energy input was $5 \times 10^{50}$ erg s$^{-1}$ (per
jet) for 10 s declining to 10$^{47}$ erg s$^{-1}$ at 1000 s.  The
initial ratio of internal energy to kinetic energy in the jet was 20
and the opening angle, 20 degrees (which was quickly reduced by 
hydrodynamical focusing). The picture shows radial velocity
80 s after the initiation of the jet.}
\label{jetsn}
\end{figure}

\subsection{MHD jets and explosions}
 
Another supernova model with us for over 30 years invokes powerful
bi-polar outflows energized by magnetic wind up and instabilities in a
differentially rotating proto-neutron star
\cite{Leb70,Bis70,Mei76}. Generically these outflows are referred to
as LeBlanc-Wilson jets. Their creation again requires very large
magnetic fields and rotation rates, once regarded as
unrealistic. Interest in this variety of model has been rekindled
however \cite{Whe00,Whe02,Ard01}, both by the observation of
jets in GRBs and by promising models for soft gamma-ray repeaters and
anomalous x-ray pulsars that invoke magnetic fields up to 10$^{15}$ G
\cite{Tho95,Tho96}.

Granted that such neutron stars exist and may be born rotating
rapidly, a robust supernova model does not necessarily follow.  A jet
is not a particularly efficient way to explode a star. Even one
introduced with a significant opening angle is rapidly collimated by
its passage through the star \cite{Zha02a} and collides with only a
small fraction of the mass (Fig. 1). Lateral shocks move around the
star, but a lot of the matter falls back, enough that it may be
difficult to preserve the neutron star. The jet may also produce very
little $^{56}$Ni, not enough to explain the light curves of Type Ib
and Ic supernovae. The velocities of the resulting supernova will be
highly asymmetric with very high values along the axis. This will give
great variation in the properties of ordinary supernovae seen at different
angles. Such variations are not seen.

This is not to say that a model for supernovae in which rotation and
magnetic fields play a major role is ruled out. The magnetic torque on
a spinning protoneutron star, $\tau = dL/dt$ with $L$ the angular
momentum, is approximately $B_r B_{\phi} R^3$, suggesting that an
angular momentum of $I \omega \sim 10^{48} (I/10^{45})(\omega/10^3)$
erg s could be braked in a few seconds if the wound up poloidal field,
$B_{\phi}$, {\sl and} radial field, $B_r$ exceeded 10$^{15}$
gauss. This would lead to the rapid dissipation of $\sim10^{51}$ erg,
possibly by Alfven waves ($\sim r^2 (\delta B)^2 v_A$ with $v_A
\sim10^{10}$ cm s$^{-1}$, the Alfven speed), long wavelength
electromagnetic waves \cite{Uso92}, or magnetic reconnection. Larger
rotation rates and stronger fields could provide greater
energies. Neutrino energy deposition and the overturn it causes might
aid in producing the necessary $B_r$.  Further work is needed here,
especially on the idea that neutrino energy deposition and MHD models
for supernovae are not exclusive.

\section{Models with Black Holes}  

Models for supernovae in which a large part of the energy comes from
an accreting black hole are newcomers to the scene, motivated
chiefly by a need to explain GRBs. However, it is recognized that
these same models may have broader applicability and, in less extreme
versions or in stars that still retain their hydrogen envelope, might
power supernovae. Such supernovae would probably retain unusual
properties such as gross asymmetry or high energy.

\subsection{Supranovae}

It has been suggested by Vietri \& Stella \cite{Vie98,Vie99} and others
that GRBs might result from the delayed implosion of rapidly rotating
neutron stars to black holes. The neutron star forms in a traditional
(neutrino-powered) supernova, but is ``supramassive'' in the sense
that without rotation, it would collapse, but with rapid rotation,
collapse is delayed until angular momentum is lost.  The momentum can
be lost by gravitational radiation and by magnetic field torques.
Vietri and Stella assume that the usual pulsar formula holds and, for
a field of 10$^{12}$ gauss, a delay of order years (depending on the
field and mass) is expected, but other parameters might give a shorter
delay. When the centrifugal support becomes sufficiently weak, the
star experiences a period of runaway deformation and gravitational
radiation before collapsing into a black hole. It is assumed that
$\sim0.1 \Msun$ is left behind in a disk which accretes and powers the
burst explosion.

As a GRB model, the supranova has several advantages. It, as well as
the collapsar model discussed later, predicts an association of GRBs
with massive stars and supernovae. Moreover it produces a large amount
of material enriched in heavy elements located sufficiently far from
the GRB as not to obscure it. The irradiation of this material by the
burst or afterglow can produce x-ray emission lines as have been
reported in several bursts \cite{Pir99,Pir00,Ree02}.  However, the
supranova model also has some difficulties \cite{Mcl02}. It may also
take fine tuning to produce a GRB days to years after the neutron star
is born. Shapiro \cite{Sha00} has shown that neutron stars requiring
differential rotation for their support will collapse in only a few
minutes.  The requirement of rigid rotation reduces the range of
masses that can be supported by rotation to, at most, $\sim$20\% above
the non-rotating limit \cite{Sha00,Sal94}.

\subsection{Collapsars}

\subsubsection{Basic collapsars}

Generically, a collapsar is a rotating massive star whose central core
collapses to a black hole surrounded by an accretion disk
\cite{Woo93,Mac99}. Accretion of at least a solar mass through this
disk produces outflows that are further collimated by passage through
the stellar mantle. These flows attain high Lorentz factor as they
emerge from the stellar surface and, after traversing many stellar
radii, produce a GRB and its afterglows by internal and external
shocks. The passage of the jet through the star also gives a very
asymmetric supernova of order 10$^{51}$ erg \cite{Zha02a}.

There are three ways to make a collapsar and each is likely
to have different observational characteristics.

\begin{itemize}

\item 
{A standard (Type I) collapsar is one where the black hole forms
promptly in a helium core of approximately 15 to 40 $\Msun$. There
never is a successful outgoing shock after the iron core first
collapses. A massive, hot proto-neutron star briefly forms and
radiates neutrinos, but the neutrino flux is inadequate to halt the
accretion.  Such an occurrence seems likely in helium cores of mass
over $\sim15 \Msun$ because of their large binding energy \cite{WHW02}
and the rapid accretion that characterizes the first second after core
collapse \cite{Fry99}.}

\item
{A variation on this theme is the ``Type II collapsar'' wherein the
black hole forms after some delay - typically a minute to an hour,
owing to the fallback of material that initially moves outwards, but
fails to achieve escape velocity \cite{Mac01}. Such an occurrence is
again favored by massive helium cores. Unfortunately the long time
scale associated with the fall back may be, on the average, too long
for typical long, soft bursts. Their accretion disks are also not
hot enough to be neutrino dominated and this may affect the accretion
efficiency \cite{Nar01} and therefore the energy available to make jets.}

\item
{A third variety of collapsar occurs for extremely massive
metal-deficient stars that probably existed only in the early universe
\cite{Abe02,Fry01}. For non-rotating stars with helium core masses
above 133 $\Msun$ (main sequence mass 260 $\Msun$), it is known that a
black hole forms after the pair instability is encountered
\cite{Heg02c}. It is widely suspected that such massive stars existed
in abundance in the first generation after the Big Bang at red shifts
$\sim$5 - 20. For {\sl rotating} stars the mass limit for black hole
formation will be raised. The black hole that forms here, about 100
$\Msun$, is more massive, than the several $\Msun$ characteristic of
Type I and II collapsars, but the accretion rate is also much higher,
$\sim$10 $\Msun$ s$^{-1}$, and the energy released may also be much
greater. The time scale is also much longer.}

\end{itemize}

For both Type I and II collapsars it is also essential that the star
loses its hydrogen envelope before death. No jet can penetrate the
envelope in less than the light crossing time, typically 100 s for a
blue supergiant and 1000 s for a red one. After running into
1/$\Gamma$ of its rest mass, a ballistic jet loses its energy.

Because of space limitations, we will not review details of
the collapsar model here, but refer the reader to the published
literature especially \cite{Mac99,Zha02a,Zha02b}. We will emphasize
however two recent developments of great interest: nucleosynthesis in
collapsar disks and the prediction by the collapsar model of other
forms of high energy transients, especially cosmological x-ray flashes
and events like GRB 980425/SN 1998bw.

\subsubsection{$^{56}$Ni production and the $r$-process}

Lacking a hydrogen envelope, the supernova that accompanies a GRB made
by a collapsar will be Type Ib or Ic with an optical luminosity given
entirely by the yield of $^{56}$Ni.  In Type I collapsars however, the
material that would have become $^{56}$Ni falls into the black
hole. The jet itself subtends a small solid angle and carries a small,
albeit very energetic mass. It cannot propagate outwards until the
mass flux inwards at the pole has declined, i.e., the density has gone
down. This makes it hard for the jet itself to synthesize much
$^{56}$Ni. How then is the supernova visible?

It is believed that the $^{56}$Ni in collapsars is made not by the
jet, but by the disk wind \cite{Mac99,Nar01}.  In the parlance of
Narayan et al., it could be that at late times (after $\sim$10 s), a
neutrino-dominated accretion disk (NDAF) switches to a convection
dominated accretion disk (CDAF) with a large fraction of the mass flow
being ejected. MacFadyen and Woosley even found considerable mass
outflow from NDAFs.  We postulate that a certain fraction of the
accreting matter - composed initially of nucleons or iron group
elements - is ejected at high velocity ($\sim$0.1 c) by the accretion
disk.

But will the material be $^{56}$Ni? Nucleosynthesis in collapsar disks
has been explored recently by Pruet and colleagues at LLNL
\cite{Pru02}.  They find that the composition flowing out from the
disk and in the jet is very sensitive to both the accretion rate and
assumed viscosity of the disk. For an ``$\alpha$-disk'' with $\alpha
\approx 0.1$ or less and accretion rates 0.1 $\Msun$ s$^{-1}$ and more the
composition will not be $^{56}$Ni, but more neutron-rich isotopes of
iron, or even $r$-process nuclei. For accretion rates around 0.01
$\Msun$ s$^{-1}$ the composition will be {\sl proton}-rich ($Y_e
\approx 0.51$), though still dominated by $^{56}$Ni. Interestingly
typical accretion rates for Type I collapsars are $\sim0.05$ $\Msun$
s$^{-1}$ (less at later times) and $^{56}$Ni synthesis is
possible. For Type II collapsars the accretion rate is lower and the
disk is proton-rich. Lower values of $\alpha$ shift the
nucleosynthesis to low $Y_e$ and for $\alpha$ = 0.01 or less,
Type I collapsar disks make no $^{56}$Ni.

\begin{figure}[t]
\begin{center}
\includegraphics[width=1.0\textwidth]{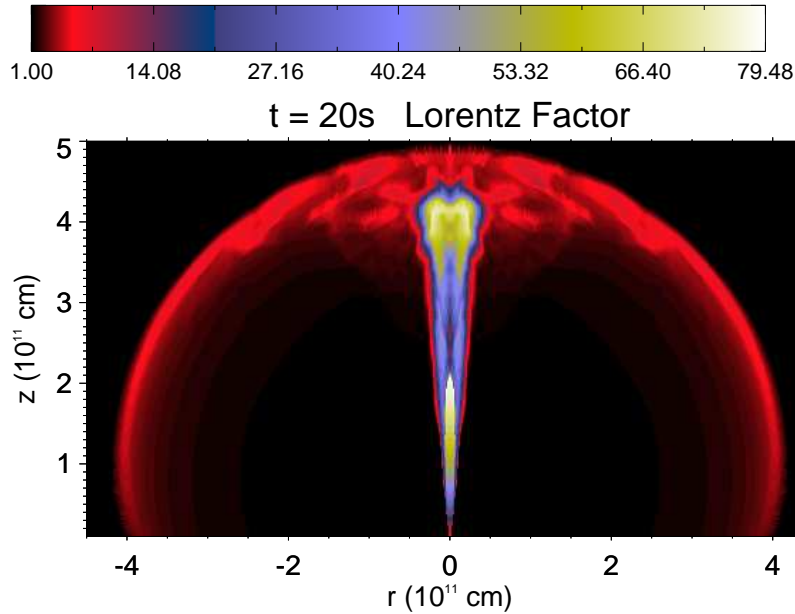}
\end{center}
\caption[]{The break out of a relativistic jet and its cocoon 22
 seconds after the jet's initiation in the star \cite{Zha02b}.}
\label{breakout}
\end{figure}

\subsubsection{X-ray flashes and supernovae}
\label{xrf}

The collapsar model was originally intended as an explanation for GRBs
but time, additional calculations, and observations suggest it
has broader implications. These essentially hinge on the answer to the
question ``If a GRB from a collapsar is only seen by observers in
about 0.3\% of the sky, what do other observers see?'' Clearly these
will be the most common events. Additionally, one may inquire what
happens when a collapsar occurs in a star still having a hydrogen
envelope \cite{Mac01}, if the parameters are such that high Lorentz
factor is not achieved, or the jet engine turns off before the jet
emerges from the star \cite{Mac00}.

In the equatorial plane of a collapsar - the common case - probably
little more is seen than an extraordinary supernova. In fact the
supernova may not even appear exceptionally energetic because the high
velocities are all along the rotational axis (Fig. 1).  Off axis
though, in a collapsar that made a GRB, one will see x-ray flashes
made by the explosion of the jet cocoon as it breaks out of the star
\cite{Ram02,Zha02b}. The cocoon contains about 10$^{50}$ - 10$^{51}$
erg \cite{Zha02a} and has Lorentz factor $\Gamma \sim 5 - 10$
(Fig. 2). By way of an external shock with the pre-explosive wind of
the stellar progenitor, this material can produce a bight transient
visible out to $\sim$30 degrees from each axis. Even though it has
lower energy per solid angle than the GRB jet (which is concentrated
within about 5 degrees), relativistic beaming compensates to make the
observable fluence comparable. That is, the GRB beams its emission to
$1/\Gamma \sim 0.005 \ {\rm radians} =$ 1/4 degree while the x-ray
flash (XRF) is beamed to perhaps 10 degrees. The duration of such
events depends on the Lorentz factor and and the pre-explosive mass
loss, but could be from tens of seconds to minutes.

These properties mesh well with the recently discovered class of
cosmological XRFs \cite{Hei01,Kip02} which share many of the
properties of long-duration GRBs (duration, frequency of occurrence,
isotropy on the sky, non-thermal spectrum, non-recurring), but have no
hard emission above about 10 keV. If our speculations are correct,
every (long-soft) GRB should have an underlying XRF that may even be
visible as a precursor to the GRB. We also predict supernovae in
association with XRFs and these might be looked for \cite{Sod02}.

\vspace{0.5\baselineskip} {\small\textbf{Acknowledgments.}  This
research has been supported, at UCSC, by the NSF (AST 02-06111), NASA
(NAG5-12036, MIT-292701), and the SciDAC Program of the DOE
(DE-FC02-01ER41176). At Chicago, work was supported by the DOE ASCI
Program (B347885).  AH is supported in part by the Department of
Energy under grant B341495 to the Center for Astrophysical
Thermonuclear Flashes at the University of Chicago.}

\end{document}